
\input harvmac.tex
\overfullrule=0pt

\def\simge{\mathrel{%
   \rlap{\raise 0.511ex \hbox{$>$}}{\lower 0.511ex \hbox{$\sim$}}}}
\def\simle{\mathrel{
   \rlap{\raise 0.511ex \hbox{$<$}}{\lower 0.511ex \hbox{$\sim$}}}}

\def\slashchar#1{\setbox0=\hbox{$#1$}           
   \dimen0=\wd0                                 
   \setbox1=\hbox{/} \dimen1=\wd1               
   \ifdim\dimen0>\dimen1                        
      \rlap{\hbox to \dimen0{\hfil/\hfil}}      
      #1                                        
   \else                                        
      \rlap{\hbox to \dimen1{\hfil$#1$\hfil}}   
      /                                         
   \fi}                                         %
\def\CL{{\cal L}}
\def\CM{{\cal M}}
\def\CO{{\cal O}}
\def\ts{\thinspace}
\def\ra{\rightarrow}

\def\Lra{\Longrightarrow}
\def\ra{\rightarrow}
\def\ol{\overline}

\def\gev{{\rm GeV}}
\def\tev{{\rm TeV}}

\def\pb{{\rm pb}}

\def\fb{{\rm fb}}
\def\ecm{\sqrt{s}}

\def\shat{\hat s}

\def\fb{{\rm fb}}

\def\Mh{M_{\eta_T}}
\def\Mv{M_{V_8}}

\def\ttb{t \ol t}
\def\bbb{b \ol b}

\def\qqb{q \ol q}

\def\ppb{p \ol p}
\def\stt{\sigma(\ttb)}
\def\Mtt{\CM_{t \ol t}}
\def\MMtt{\langle \Mtt \rangle}

\def\RMStt{{\langle \Mtt^2 \rangle^{1/2}}}
\def\Deltt{\Delta \Mtt}

\def\jets{{\rm jets}}
\def\pt{p_T}
\def\et{E_T}

\def\Dzero{{\rm D}\slashchar{\rm O}}
\def\et{E_T}
\def\etmiss{\slashchar{E}_T}
\def\cstar{\cos \theta^*}

\def\hl{10^{33} \ts {\rm cm}^{-2} \ts {\rm s}^{-1}}

\def\myfoot#1#2{{\baselineskip=14.4pt plus 0.3pt\footnote{#1}{#2}}}

\Title{\vbox{\baselineskip12pt\hbox{BUHEP--95--2}}}
{Top Quarks and Flavor Physics}

\bigskip
\centerline{Kenneth Lane\myfoot{$^{\dag }$}{lane@buphyc.bu.edu}}
\smallskip\centerline{Department of Physics, Boston University}
\centerline{590 Commonwealth Avenue, Boston, MA 02215}
\vskip .3in

\centerline{\bf Abstract}

Because of the top quark's very large mass, about 175~GeV, it now provides
the best window into flavor physics. Thus, pair--production of top quarks
at the Tevatron Collider is the best probe of this physics until the Large
Hadron Collider turns on in the next century. I discuss aspects of the
mass and angular distributions that can be measured in $\ttb$ production
with the coming large data samples from the Tevatron and even larger ones
from the LHC.

\bigskip

\Date{1/95}

\vfil\eject

\newsec{Introduction}

The CDF collaboration has reported evidence for top--quark production at
the Tevatron Collider
\ref\cdfpr{F.~Abe, et al., The CDF Collaboration, Phys.~Rev.~Lett.~{\bf
73}, 225 (1994); Phys.~Rev.~{\bf D50}, 2966 (1994).}.
According to these
papers, the top mass is $m_t = 174\pm 10 \ts ^{+13}_{-12} \,\gev$. The CDF
data are based on an integrated luminosity of $19.3\,\pb^{-1}$.
Taking into account detector efficiencies and acceptances, CDF reports
the production cross section $\sigma(\ppb \ra \ttb) = 13.9
\ts^{+6.1}_{-4.8}\,\pb$ at $\sqrt{s} = 1800\,\gev$. The predicted QCD cross
section for $m_t = 174\,\gev$, including next--to--leading--log
corrections
\ref\qcdref{P.~Nason, S.~Dawson, and R.~K.~Ellis, Nucl.~Phys. {\bf B303}
(1988)~607; W.~Beenakker, H.~Kuijf, W.~L.~van~Neerven and
J.~Smith, Phys.~Rev. {\bf D40} (1989)~54.},
and soft--gluon resummation
\ref\resum{E.~Laenen, J.~Smith and W.~L.~Van Neerven, Nucl.~Phys.~{\bf B369}
(1992)~543; {\it ibid}, FERMILAB--Pub--93/270--T.},
is $\stt = 5.10^{+0.73}_{-0.43}\,\pb$. This is 2.8~times smaller than the
central value of the measured cross section. The uncertainty in $\alpha_S$
increases the theoretical error in $\stt$ to at most~30\%
\ref\ellis{K.~Ellis, ``Top--Quark Production Rates in the Standard
Model'', invited talk at the 27th International
Conference on High Energy Physics, Glasgow, 20--27th July 1994.}.

Very recently, the $\Dzero$ Collaboration has also reported evidence for
top--quark production
\ref\dz{S.~Abachi, et al., The $\Dzero$ Collaboration, ``Search for High
Mass Top Quark Production in $\ppb$ Collisions at $\ecm = 1.8\,\tev$'',
hep-ex--9411001, submitted to Physical Review Letters, November 9, 1994.}.
A direct measurement of the top--quark mass and cross section was not made
by $\Dzero$ in this report. However, assuming that the excess of signal
events over expected background is due to $\ttb$ production and that $m_t =
180\,\gev$, $\Dzero$ deduces the cross section $\stt = 8.2\pm 5.1\,\pb$.
This is consistent with the standard model and with CDF.

The experimental errors on the CDF and $\Dzero$ measurements are large.
Assuming the top mass is close to $175\,\gev$, the CDF cross section
could be due to an up--fluctuation or to underestimated
efficiencies (although the latter seems unlikely; see~\cdfpr).
But, if it is confirmed by both experiments in their current
higher--luminosity runs, the large $\ttb$ rate heralds the long--awaited
collapse of the standard model. Even if the standard model result is found
in the new data, however, it is clear that the top quark provides a
wide--open window into the world of flavor physics. It is the heaviest
elementary particle we know and, more to the point, the heaviest elementary
fermion by a factor of~40---as massive as tungsten. As a first example of
flavor physics, we note that if the Higgs boson of the minimal one--doublet
model exists, its coupling to the top quark, renormalized
at~$m_t=174\,\gev$, is large: $\Gamma_t = 2^{3/4}\ts G_F^{1/2}\ts m_t =
1.00$. If there are charged scalars, members of Higgs--boson multiplets or
technipions, they are expected to couple to top quarks with $\CO(1)$ strength
and to decay as $H^+ \ra t \ol b$. Recently, several papers have discussed
aspects of flavor physics that are more or less intimately connected to the
large top mass and that lead to enhanced rates of the $\ttb$ signals
studied by the Tevatron experiments
\ref\hp{C.~Hill and S.~Parke, Phys.~Rev.~{\bf D49}, 4454 (1994)},
\ref\eekleta{E.~Eichten and K.~Lane, Phys.~Lett.~{\bf B327}, 129 (1994).},
\ref\bp{V.~Barger and R.J.N.~Phillips, U.~Wisconsin Preprint,
MAD/PH/830 (May 1994).}.

In this paper, we discuss two distributions that may be used to distinguish
alternative models of $\ttb$ production---including standard QCD
\ref\ichep{An abridged version of this work was given in
K.~Lane, ``Top--Quark Production and Flavor Physics---The Talk'',
to appear in the Proceedings of the 27th International Conference on High
Energy Physics, Glasgow, 20--27th~July 1994; Boston University Preprint
BUHEP--94--25 (1994).}.
These are the invariant mass distribution, $d\sigma/d\Mtt$, and the
center--of--mass (c.m.) angular distribution of the top quark,
$d\sigma/d\cos\theta$. The magnitude and shape of the invariant mass
distribution (Section~2) can reveal whether $\ttb$ production is standard
or not, and whether resonances decaying to $\ttb$ exist. We point out that,
for standard QCD production, the mean and root--mean--square invariant
masses are linear functions of the top--quark mass over the entire
interesting range of $m_t$. Thus, the $\Mtt$ distribution can provide an
{\it independent} determination of the top quark's mass. We apply this to
the CDF data~\cdfpr\ and find quite good consistency with the directly
measured mass. This analysis is made at the most elementary theoretical
level. It needs to be carefully redone by the experimental
collaborations themselves.

In Section~3, we apply the $\Mtt$ analysis to examples
of the three nonstandard mechanisms of $\ttb$ production described in
Refs.~\hp, \eekleta\ and \bp. The first involves
resonant production of a 400--600~GeV color--octet vector meson
(``coloron''), $V_8$, which is associated with electroweak symmetry
breaking via top--condensation
\ref\topcolor{C.~T.~Hill, Phys.~Lett.~{\bf 266B}, 419 (1991)\semi
S.~P.~Martin, Phys.~Rev.~{\bf D45}, 4283 (1992);
{\it ibid}{\bf D46}, 2197 (1992).}
and which interferes with
QCD production via the process $\qqb \ra V_8 \ra \ttb$.
The second example invokes a color--octet pseudoscalar, $\eta_T$
\ref\etatrefs{E.~Farhi and L.~Susskind Phys.~Rev.~{\bf D20}
(1979)~3404\semi
S.~Dimopoulos, Nucl.~Phys.~{\bf B168} (1980)~69 \semi
T.~Appelquist and G.~Triantaphyllou, Phys.~Rev.~Lett.~{\bf
69},2750 (1992) \semi
T.~Appelquist and J.~Terning, Phys.~Rev.~{\bf D50}, 2116 (1994).}.
In multiscale models of walking technicolor
\ref\multi{K.~Lane and E.~Eichten, Phys.~Lett.~{\bf 222B} (1989)~274 \semi
K.~Lane and M~V.~Ramana, Phys.~Rev.~{\bf D44} (1991)~2678.},
\ref\wtc{B.~Holdom, Phys.~Rev.~{\bf D24} (1981)~1441;
 Phys.~Lett.~{\bf 150B} (1985)~301 \semi
T.~Appelquist, D.~Karabali and L.~C.~R. Wijewardhana,
Phys.~Rev.~Lett.~{\bf 57} (1986) 957~;
T.~Appelquist and L.~C.~R.~Wijewardhana, Phys.~Rev.~{\bf D36} (1987)~568
\semi  K.~Yamawaki, M.~Bando and K.~Matumoto, Phys.~Rev.~Lett.~{\bf 56},
(1986)~1335 \semi
T.~Akiba and T.~Yanagida, Phys.~Lett.~{\bf 169B} (1986)~432.},
the $\eta_T$ is produced strongly in gluon--gluon fusion and decays mainly
to $\ttb$. The third model has additional production of the classic $\ttb$
signals~\cdfpr. This occurs through pair--production of an
electroweak--{\it isoscalar}, color--triplet quark, $t_s$, which is
approximately degenerate with the top quark and which, through
mass--mixing, decays as $t_s \ra W^+ b$. The agreement found in Section~2
between CDF's directly measured top--mass and that extracted from the
$\Mtt$ moments does not yet rule out these new mechanisms of top--quark
production. The $\Mtt$ distributions from the current Tevatron run may do
so. (Of course, finding the standard model cross section will also be
powerful evidence against alternative production mechanisms.)

The angular distribution of top quarks (Section~4) also reflects the
underlying production mechanism. Even though most of $\ttb$ production is
near threshold, the expectation that it is mainly s--wave can be overturned
if there are large parity--violating components in the $\qqb \ra \ttb$
process. We shall compare the angular distributions for standard and
nonstandard $\ttb$ production at the Tevatron and at the CERN Large Hadron
Collider. We shall see that, because of the much larger $\tau =  \shat/s$
in top--pair production at the Tevatron, experiments there have an
advantage over those at the LHC.\foot{In this paper I do not discuss
high--energy $e^+e^-$ colliders such as the 500~GeV (or so) NLC.  Our focus
is on distinguishing alternate mechanisms of $\ttb$ production. Lepton
machines cast no light on such strongly--coupled flavor physics aspects of
$\ttb$ production as the $V_8$ and $\eta_T$. The higher rates possible at
hadron machines also make them ideal for searches for new particles such as
charged scalars in the decays of top quarks.} These angular tests require
much larger data sets than will be available in the next year or two. To
realize the full potential of this handle on flavor physics, it is
essential that the Tevatron experiments be able to collect samples as large
as 1--10~fb$^{-1}$.

\newsec{Invariant Mass Distributions In QCD}

Strictly speaking, the $\ttb$ invariant mass $\Mtt$ is not well--defined in
QCD because of the emission of soft and collinear gluons from the $t$ and
$\ol t$~quarks. Nevertheless, the theoretical
invariant mass is numerically not very different from a definition of
$\Mtt$ that allows for this gluon radiation. Furthermore, because the $\pt$
of the $\ttb$ c.m.~system typically is small compared to $m_t$, it will be a
good approximation for us to use the lowest--order QCD cross section
$d\sigma/d\Mtt$ to discuss the moments of the invariant mass distribution.
This distribution is shown in Fig.~1 for the Tevatron Collider and for
top--quark masses in the interesting range 100--220~GeV.%
\foot{These plots and all other calculations in this paper
were carried out using lowest--order QCD subprocess cross sections and the
EHLQ Set~1 parton distribution functions
\ref\ehlq{E.~Eichten, I.~Hinchliffe, K.~Lane and C.~Quigg, Rev.~Mod.~Phys.
{\bf 56}, 579 (1984).}.
To account for QCD radiative corrections, our $\ttb$ cross sections have
been multiplied by 1.62. This makes our QCD rates and the central values
quoted in Ref.~\resum\ agree to within one per~cent over the entire
interesting range of top masses. Our numerical results for the linear
dependence of $\MMtt$ and $\RMStt$ on $m_t$ are accurate so long as the
higher--order corrections are well--represented by a simple multiplicative
factor.}
The mass distributions are seen to be sharply peaked at ${\cal M}_{\rm max}
\simeq 2.1 m_t + 10\,\gev$. Consequently, low moments of the mass
distribution, the mean and RMS, are nearly linear functions of the
top--quark mass (also see Ref.~\ref\tdr{{\it GEM Technical Design Report},
Chapter~2; GEM TN--93--262, SSCL--SR--1219; Submitted by the GEM
Collaboration to the Superconducting Super Collider Laboratory (April 30,
1993)\semi K.~Lane, F.~Paige, T.~Skwarnicki and J.~Womersley, {\it
Simulations of Supercollider Physics}, hep-ph--9412280, submitted to
Physics Reports.}). For $100 \simle m_t \simle 200\,\gev$,
the first two moments are well--fit by the formulae
\eqn\mttvsmt{\eqalign{
\MMtt &= 50.0\,\gev + 2.24 \ts m_t \cr
\RMStt &= 58.4\,\gev + 2.23 \ts m_t \ts. \cr }}
In the range $m_t \simeq$ 140--200~GeV, the dispersion in $\Mtt$ expected
for standard QCD production is $\Deltt =$ 70--80~GeV.%
\foot{If there are experimental difficulties in measuring $\Mtt$ that do
not also affect the measurement of $m_t$, one could instead use the mean
value of the summed scalar--$\et$ to extract the top--quark mass. Indeed, in
Ref.~\tdr, it was shown that a quantity as indirect as the invariant mass
$\CM_{e\mu}$ of the isolated electron plus muon measured in $\ttb \ra e^\pm
+ \mu^\mp + \jets$ is also a linear function of $m_t$ and may be used to
determine it.}

In Ref.~\cdfpr, the top quark mass was determined from a sample of seven $W
\ra \ell \nu \ts + \ts 4 \ts \ts \jets$ events by making an overall
constrained best fit to the hypothesis $\ppb \ra \ttb + X$, followed by the
standard top decays $t \ra W^+ b$ with one $W$ decaying leptonically and
the other hadronically. At least one of the $b$--jets was tagged. The CDF
paper provides the momentum 4--vectors of all particles in the event before
and after the constrained fit. From these, the central values of kinematic
characteristics of the seven events may be determined. Table~1 lists the
best--fit top--quark masses determined by CDF together with the invariant
mass of the events before and after the constrained fit.\foot{Particle
4--vectors before the constrained fit do have various corrections---e.g.,
for the jet energy scale---made to them~\cdfpr. Only $\etmiss$ is provided
for the neutrino(s) in the before--fit 4--vectors. The biggest change in
the before and after momenta occurs in $\etmiss$. We used the $W \ra
\ell \nu$ 4--momenta determined from the constrained fit in both cases.} We
used these $\Mtt$ to compute the mean and RMS. Both sets of 4--momenta gave
essentially identical results. Using 4--momenta from the constrained fit,
we found:
\eqn\after{\eqalign{
\MMtt &= 439 \pm 11\,\gev \quad \Lra \quad m_t = 173 \pm 5\,\gev \cr
\RMStt &= 443 \pm 11\,\gev \quad \Lra \quad m_t = 172 \pm 5\,\gev \cr
\Deltt &= 59.5\,\gev \ts. \cr}}
The errors in Eq.~\after\ were estimated by the ``jacknife'' method of
computing the moments while omitting one of the seven events. They give some
sense of the theoretical error in determining the mean and RMS invariant
masses from the limited CDF sample. They are {\it not} to be interpreted as
the true experimental errors; the CDF group must provide those. However, we
expect that the process of averaging the invariant mass will give
moderately small experimental errors.

These results give some confidence that CDF's measured central value of the
top--quark mass, 174~GeV, is accurate. For example, if $m_t = 160\,\gev$
(for which Ref.~\resum\ predicts $\stt = 8.2^{+1.3}_{-0.8}\,\pb$), we would
expect $\MMtt = 409\,\gev$ and $\RMStt = 415\,\gev$, well below the
values determined above. Thus, if something is going to change in the CDF
results from the current run, we expect it will be the cross
section---which needs to become two to three times smaller to agree with
the standard model.

\newsec{Nonstandard Mass Distributions}

In this section we examine three nonstandard proposals~\hp, \eekleta, \bp\
for $\ttb$ production and the large cross section reported by CDF~\cdfpr.
We shall find that they are not yet disfavored by the good agreement
between the central values of the measured top mass and the top masses
deduced in Eq.~\after. We begin by quoting the differential cross sections
for $\qqb \ra \ttb$ and $gg \ra \ttb$ in lowest--order QCD:
\eqn\qcdang{\eqalign{
{d \hat \sigma(\qqb \ra \ttb) \over {d z}} =
{\pi \alpha_s^2 \beta \over {9 \shat}} \ts &\bigl(2 - \beta^2 + \beta^2
z^2\bigr) \ts,
\cr\cr
{d \hat \sigma(gg \ra \ttb) \over {d z}} =
{\pi \alpha_s^2 \beta \over {6 \shat}} \ts
&\biggl\{{1 + \beta^2 z^2 \over {1 - \beta^2 z^2}} -
{(1-\beta^2)^2 \ts (1 + \beta^2 z^2) \over{(1-\beta^2 z^2)^2}}
- \textstyle{{9\over{16}}} (1 + \beta^2 z^2) \cr
& + {1-\beta^2 \over{1-\beta^2 z^2}} \ts (1 -
\textstyle{{1\over{8}}} \beta^2 + \textstyle{{3\over{8}}} \beta^2 z^2)
\biggr\} \ts, \cr}}
where $z = \cos\theta$, $\theta$ is the c.m.~scattering variable and $\beta
= \sqrt{1 - 4 m_t^2/\shat}$. For $\shat \gg 4 m_t^2$, these cross
sections---especially the gluon fusion one---are forward--backward peaked.
But, at the modest $\shat$ at which QCD $\ttb$ production is large, the
cross sections are fairly isotropic.

For the ``coloron'' bosons of Ref.~\hp, we adopted a version of the model
in which the gauge symmetry $SU(3)_1 \otimes SU(3)_2$ breaks down to color
$SU(3)$, yielding eight massless gluons and equal-mass $V_8$'s. To study
parity violation in the angular distributions in $\ttb$ production
(see Section~4), we made the theoretically inane assumption that the
$V_8$ couples only to left--handed quarks with the amplitude
\eqn\topvqq{
A(V^a_8(p,\lambda) \ra q(p_1) \ts \ol q(p_2)) =
g_s \ts \xi_q \ts \epsilon^\mu(p,\lambda) \ts
\ol u_q(p_1) \ts {\lambda_a \over{2}} \ts \gamma_\mu \ts \left({1-\gamma_5
\over{2}}\right)\ts v_q (p_2) \ts.}
Here, $g_s$ is the QCD coupling and, following Ref.~\hp, we took $\xi_t =
\xi_b = \pm 1/\xi_q$ ($q =u,d,c,s$). For this chiral coupling, the $\qqb
\ra \ttb$ angular distribution in Eq.~\qcdang\ is modified by the addition of
\eqn\sigveight{
{d \hat \sigma(\qqb \ra V_8 \ra \ttb) \over {d z}} =
{\pi \alpha_s^2 \beta \over {36 \shat}} \ts (1 + \beta z)^2
\ts \left\{\ts\biggl|1 + \xi_q \ts\xi_t \ts{\shat \over {\shat - \Mv^2 + i
\sqrt{\shat} \ts \Gamma(V_8)}}\biggr|^2 - 1\right\} \ts,}
where, ignoring the mass of all quarks except the top's, the $V_8$ width is
\eqn\widveight{
\Gamma(V_8) = {\alpha_s \Mv \over {12}}\ts
\biggl\{4\xi_q^2 + \xi_t^2 \left(1 + \beta_t (1-m_t^2/\Mv^2)\right)\biggr\}
\ts,}
where $\beta_t = \sqrt{1 - 4 m_t^2/\Mv^2}$.

The $\Mtt$ distribution in the coloron model for $\Mv = 450\,\gev$ is shown
in Fig.~3 for $\xi_t = \xi_b = - 1/\xi_q = \sqrt{40/3}$ (see \hp) and in
Fig.~4 for $\xi_t = \xi_b = 1/\xi_q = \sqrt{40/3}$. The effect of
interference with the QCD amplitude is obvious as is the tendency for the
$\Mtt$ distribution to be enhanced at lower (higher) masses for $\xi_t =
-1/\xi_q$ ($+1/\xi_q$). The theoretical width of the $V_8$ in this example
is $\Gamma(V_8) \cong \Gamma(V_8 \ra \bbb) + \Gamma(V_8 \ra \ttb) =
40\,\gev$. Figures~5 and 6 show the mass distibutions for $\Mv = 475\,\gev$
and $\xi_t = \mp\xi_b = \sqrt{40/3}$. Here, $\Gamma(V_8) \cong 85\,\gev$
and the mass distribution is significantly broader than in the case of $\Mv
= 450\,\gev$. The characteristics of these mass distributions will be
discussed below together with those of the other nonstandard production
models we are considering.\foot{Note that this $V_8$ model predicts a
strong resonance in $\qqb \ra \bbb$, providing another good way to to test
it.}

Many technicolor models contain a color--octet pseudoscalar boson,
$\eta_T$. So long as the $\eta_T$ may be treated as a pseudo--Goldstone
boson, its decay rates to gluons can be computed from the triangle
anomaly~\etatrefs. We introduce a dimensionless factor $C_q$ in the Yukawa
coupling of $\eta_T$ to $\qqb$~\eekleta. While it is determined by the
details of the underlying extended technicolor model, we expect $C_q =
\CO(1)$. Then, the $\eta_T$'s main decay modes are to two gluons and $\ttb$
and they are given by
\eqn\etarates{\eqalign{
\Gamma(\eta_T \ra gg) &= {5 \alpha_s^2 \ts N_{TC}^2 \ts \Mh^3  \over {384 \ts
\pi^3 \ts F_Q^2}} \ts, \cr\cr
\Gamma(\eta_T \ra \ttb) &= {C_t^2 \ts m_t^2 \ts \Mh \ts \beta_t \over {16 \pi
F_Q^2}} \ts.\cr}}
The gluon fusion cross section for $\ttb$ production is modified by the
addition of
\eqn\sigeta{\eqalign{
{d \hat \sigma(gg \ra \eta_T \ra \ttb) \over {d z}} =
&{\pi \over{4}} \ts {\Gamma(\eta_T \ra gg)\ts \Gamma(\eta_T \ra \ttb) \over
{(\shat - \Mh^2)^2 + \shat \ts \Gamma^2(\eta_T) }} \cr\cr
&\ts\ts + {5 \sqrt{2} \ts \alpha_s^2 \ts N_{TC}\ts C_t \ts m_t^2 \ts \beta
\over {768 \pi F_Q^2}} \ts {\shat - \Mh^2 \over
{(\shat - \Mh^2)^2 + \shat \ts \Gamma^2(\eta_T) }} \ts
\ts {1 - 2 \beta^2 z^2 \over {1 - \beta^2 z^2}} \ts. \cr}}
In these expressions, it is assumed that the $\eta_T$ is composed from a
single doublet of techniquarks $Q = (U,D)$ in the ${\bf N_{TC}}$
representation of $SU(N_{TC})$; $F_Q$ is the decay constant of technipions
in the $\ol Q Q$ sector. The first term on the right is isotropic; the
second (interference) term is never very important, but we include it for
completeness. In the narrow resonance approximation, the contribution of
the $\eta_T$ to the $p p^\pm \ra \ttb$ rate,
\eqn\narrowh{\eqalign{
\sigma(p p^\pm \ra \eta_T \ra \ttb) &\simeq {\pi^2 \over {2s}} \ts
{\Gamma(\eta_T \ra gg) \ts \Gamma(\eta_T \ra \ttb) \over {\Mh \ts
\Gamma(\eta_T)}} \cr
&\qquad \times \int d \eta_B \ts f_g^p\biggl({\Mh\over{\sqrt{s}}}
e^{\eta_B}\biggr)
\ts f_g^p\biggl({\Mh\over{\sqrt{s}}} e^{-\eta_B}\biggr) \ts,}}
scales as $N_{TC}^2/F_Q^2$. Here, $\eta_B$ is the boost rapidity of the
$\ttb$ c.m.~frame and $f_g^p(x)$ is the gluon distribution function for
the proton for mommentum fraction $x$ and $Q^2 = \Mh^2$.

As discussed in Ref.~\eekleta, the $\eta_T$ of the standard one--family
technicolor model~\etatrefs\ has $F_Q = 123\,\gev$ and, for $N_{TC} \simle
8$, it cannot significantly increase the $\ttb$ rate at the
Tevatron. Thus, we were motivated to consider the $\eta_T$ arising in
multiscale models~\multi\ of walking technicolor~\wtc. Multiscale models
are characterized by a small $\eta_T$ decay constant; for the calculations
presented here, we chose $F_Q = 30\,\gev$. The mass distribution for a
model with $N_{TC} = 5$ and $C_t = -1/3$ is shown in Fig.~7 for $\Mh =
450\,\gev$. The width of the $\eta_T$ in this case is $\Gamma(\eta_T) \cong
\Gamma(\eta_T \ra \ttb) + \Gamma(\eta_T \ra gg) = 21\,\gev + 11\,\gev =
32\,\gev$. Figure~8 shows the cross section for $\Mh = 475\,\gev$; in this
case, $\Gamma(\eta_T) \cong 37\,\gev$. In both cases, the small decay
constant results in a rate 2--3 times larger than QCD.

The third model of enhanced top--production we considered is one in which
an electroweak--isoscalar, charge ${\textstyle {2\over { 3 }}}$ quark,
$t_s$, is approximately degenerate with the top--quark and mixes with it so
that both have the same $Wb$~decay mode~\bp. If $m_{t_s} = m_t = 174\,\gev$
the expected rate for the top--quark signal is doubled to $10.2\,\pb$. We
illustrate the isoscalar quark model in Figs.~9 and 10 with two cases:
$m_{t_s} = 160$ and $m_t = 175\,\gev$; $m_{t_s} = 165$ and $m_t =
190\,\gev$.

Before discussing features of these nonstandard $\Mtt$ distributions, a
comment on radiative corrections is in order. As noted above, we have
multiplied all our lowest--order EHLQ1 cross sections by the factor 1.62.
This is a composite of the radiative corrections {\it at the Tevatron} for
the purely QCD processes $gg \ra \ttb$ and $\qqb \ra \ttb$. For a 1.8~TeV
$\ppb$ collider, the $\qqb$ process accounts for 90\% of heavy $\ttb$
production in the standard model.\foot{This situation is reversed at the
15~TeV LHC $pp$ collider, where $gg \ra \ttb$ is 90\% of the QCD cross
section.} On the other hand, the gluon fusion process receives the largest
radiative correction~\qcdref, \resum. We do not know the radiative
corrections appropriate to the resonant production processes $\qqb \ra V_8
\ra \ttb$ and $gg \ra \eta_T \ra \ttb$, but it seems likely that our
multiplication by 1.62 overestimates the former and underestimates the
latter process. Thus, the total Tevatron cross sections for these processes
may be accurate to only about~30\%.

The total $\ttb$ cross sections at the Tevatron and the characteristics
extracted from the $\Mtt$ distributions are displayed in Table~2 for the
CDF data (see Eq.~\after) and for the three nonstandard production models
described above. We note the following features:

\medskip

\noindent 1.) The CDF data is narrower ($\Deltt = 60\,\gev$) than the QCD
expectation (77~GeV). While this $\Deltt$ is consistent with the resonant
production models, the statistics are so low that we do not consider this
significant. It is a feature worth watching for in future data samples.

\noindent 2.) If $\xi_q \xi_t = -1$ in the coloron model (corresponding to
the notation $V_8^-$ in Table~2), the mass distribution is increased below
the resonance and depressed above it; vice-versa for $\xi_q \xi_t = +1$
($V_8^+$). We see that, for a given $\Mv$, this results in an extracted
value of $m_t$ that is somewhat smaller than or {\it significantly} larger
than the directly--measured one, depending on whether $\xi_q \xi_t = -1$ or
$+1$.

\noindent 3.) The $\eta_T$'s we have considered are narrow enough to not
interfere appreciably with the QCD gluon fusion process. Thus, the value of
the extracted top mass depends mainly on $\Mh$; it tends to be larger for a
larger $\Mh$, but then the $\eta_T$ rate becomes smaller and the distortion
of $\stt$ less important. Resonance masses in the range 400--500~GeV return
a top mass close to the directly--measured value.

\noindent 4.) It is easy to double the QCD value of $\stt$ in the isoscalar
quark model: just choose $m_{t_s} = m_t$. But, as could be foreseen, it is
difficult for the isoscalar quark model to give both a 13.9~pb cross
section and an extracted mass close to the directly--measured one. To get a
cross section $\sim 3$~times as large as QCD requires choosing one of the
masses significantly lower than 174~GeV, leading to too small an extracted
value. This model could be the easiest to eliminate with data from the
current Tevatron run.

\medskip

Finally, we remark that subsystem invariant masses may be as interesting as
the total invariant mass. For example, in multiscale technicolor, it is
possible that a color--octet technirho is produced and decays as $\rho_T
\ra W^\mp \pi_T^\pm $, with $\pi_T^+ \ra t \ol b \ra W^+ b \ol b$, the {\it
same} final state as in $\ttb$ production~\multi. Searches for processes
such as these, using a constrained--fit procedure analogous to that
employed by CDF for the $\ttb$ hypothesis, should be carried out. All this
will require a lot of data from the Tevatron, perhaps $1\,\fb^{-1}$ or
more. At the expense of increasing backgrounds, larger data samples may be
had by using appropriately selected events {\it without} a tagged $b$--jet.
This was done in Ref.~\cdfpr\ and was found to give an excess of events
with constrained--fit $m_t$ above 160~GeV.

\newsec{Angular Distributions}

The $\ttb$ angular distribution of top quarks also provides information
about their production mechanism. The distribution expected in
lowest--order QCD was given in Eq.~\qcdang. As we noted, the $gg \ra \ttb$
process---10\% of the QCD rate at the Tevatron and 90\% of it
at the LHC---is strongly forward--backward peaked at $\shat \gg 4~m_t^2$,
but fairly isotropic near threshold where most $\ttb$ production occurs.
Resonances such as the top--color $V_8$ and the technicolor $\eta_T$ may
change the proportion of $gg$ and $\qqb$--induced $\ttb$ production and the
top angular distribution at these colliders.

By Bose symmetry, the angular distribution in $gg \ra \ttb$ processes is
forward--backward symmetric. Although this is also true in lowest--order
QCD for $\qqb \ra \ttb$, there is no reason it need be so for nonstandard
production mechanisms. To illustrate the ability of hadron collider
experiments to distinguish different angular distributions, we assume that
the coloron $V_8$ couples only to left--handed quarks, implying the angular
distribution $(1 + \beta \cos \theta)^2$.

The Tevatron has a distinct advantage in the study of top angular
distributions. To determine $\theta$ in $\qqb \ra \ttb$ processes, we need
to know the direction of the incoming light quark as well as that of the
outgoing top quark.\foot{The distinction between $t$ and $\ol t$ is based
on the sign of the charged lepton in $W$--decay.} In $\ppb \ra \ttb$ at
$\sqrt{s} = 1800\,\gev$, the $q$--direction is the same as that of the
proton practically all the time. Thus, if we denote by $\theta^*$ the angle
between the proton direction and the top--quark direction in the
subprocess~c.m., this angle is almost always the same as $\theta$. As we
shall see, the Tevatron's analyzing power would be significantly improved
if the luminosity of the Tevatron were increased to $\hl$ or more and its
detectors upgraded to handle this luminosity.

In $pp$ collisions, the direction of the incoming quark can be inferred
with confidence only for events with high boost rapidity, $\eta_B$, or
large fractional subprocess energy, $\tau = \shat/s$. (For $pp$ collisions,
$\theta^*$ will be defined as the angle between the direction of the boost
and that of the top quark in the subprocess~c.m.) For large $\tau$, the
quark direction tends to be the same as the boost of the c.m., even if
$\eta_B$ is small~\tdr. However, $\tau$ is small for $\qqb \ra \ttb$ at the
LHC, making it hard to distinguish $\theta$ from $\pi - \theta$. To
make matters worse, $\ttb$ production is dominated by gluon fusion,
obscuring any interesting $\cos\theta$ dependence. Thus, angular
information on top production is doubly difficult to come by at the LHC.

The $\cstar$ distributions we present below are integrals over $\ttb$
invariant mass of $d \sigma(p p^\mp \ra \ttb)/d\Mtt \ts d\cstar$. The
integration region is centered on the peak of the invariant mass
distribution and is approximately the width of the resonance. For the
$\eta_T$, we used $\Mh = 450\,\gev$, $N_{TC} = 5$, $F_Q = 30\,\gev$ and
$C_t = -1/3$. Its width is 32~GeV. For the $V_8$, we took $\Mv = 475\,\gev$
and $\xi_t = \mp 1/\xi_q = \sqrt{40/3}$. The $V_8$ width is 85~GeV.

The $\cstar$ distributions, defined as described above for $\ppb$
(Tevatron) and $pp$ (LHC) collisions, are shown for the $\eta_T$ and $V_8$
models in Figs.~11--14. Global features of these distributions are
summarized in Table~3. The top quarks were required to have pseudorapidity
$|\eta| < 2$, which we estimate to correspond to the average acceptance of
the CDF and $\Dzero$ detectors for leptons and jets from top
decay.\foot{Our results do not change significantly if we require $|\eta| <
1.5$ for the Tevatron detectors and allow $|\eta| < 2.5$ for the LHC
detectors.} We discuss them in turn:

Figure 11 shows the $\qqb \ra \ttb$, $gg \ra \ttb$ and $gg \ra \eta_T \ra
\ttb$ components of the top--production $\cstar$ distribution expected at
the Tevatron. The $\Mtt$ integration region is 430~to 470~GeV. The QCD
contribution is flat, the forward--backward peaking diminished by the
proximity of threshold. The $\eta_T$ contribution is also flat, of course,
and makes up about 85\% of the total cross section. The falloff above
$|\cstar| = 0.90$ is due to the rapidity cut, $|\eta_{t, \ol t}| < 2.0$.
(We computed the $\cstar$ distribution of the seven $\ttb$~candidate events
reported by CDF~\cdfpr. The results, along with the top quark's
c.m.~velocity~$\beta$, are listed in Table~1. They form a perfectly flat
distribution.) Table~3 lists the total $\ttb$~cross section as well as the
cross sections $\sigma_F$ for $\cstar > 0$ and $\sigma_B$ for $\cstar < 0$.
The forward-backward asymmetry is calculated as
\eqn\afb{
A_{FB} = {N_F - N_B \over{N_F + N_B}} =
{\sigma_F - \sigma_B \over{\sigma_F + \sigma_B}} \ts.}
The statistical error on $A_{FB}$ is
\eqn\Dafb{
(\Delta A_{FB})_{\rm stat} = 2 \sqrt{{N_F N_B \over {(N_F + N_B)^3}}} =
2 \sqrt{{\sigma_F \sigma_B \over {(\sigma_F + \sigma_B)^3 \ts
\epsilon_{t \ol t}\ts \int \CL dt}}} \ts,}
where $\epsilon_{t \ol t}$ is the overall efficiency, including branching
ratios, for identifying and reconstructing $\ttb$ events. For the CDF
experiment at the Tevatron, we can infer from Ref.~\cdfpr\ that
$\epsilon_{t \ol t}({\rm CDF}) \simeq$ 5--10 events$/(19\,\pb^{-1} \times
14\,\pb) =$ 2--4\%. We use $\epsilon_{t \ol t}({\rm TEV}) = 3\%$. It is
difficult to say what value of the efficiency is appropriate for LHC
experiments; detailed simulations are needed (see e.g., Ref.~\tdr).
We shall assume $\epsilon_{t \ol t}({\rm LHC}) = 5\%$, although it turns
out not to matter in the examples we consider.

The components of the $\cstar$ distribution expected at the LHC are shown
in Fig.~12. Because of the small $\tau$ values involved, the roles of gluon
fusion and $\qqb$ annihilation are reversed, with gluon fusion making up
about 90\% of the QCD rate. The enormous $\eta_T \ra \ttb$ rate is due to
the very large $gg$~luminosity at small~$\tau$~\ehlq. The slight central
bowing of the $\cstar$ distribution is due to the top--rapidity cut. At the
LHC energy, such large boost rapidities occur that events at large
c.m.~rapidity and $\cstar$ are depleted.

Figure 13 shows the components of the $\cstar$ distribution at the Tevatron
for the 475~GeV $V_8$ coupling to left--handed quarks with relative
strengths $\xi_t = -1/\xi_q = \sqrt{40/3}$. The $\Mtt$ integration region
is 400--500~GeV. The effect of the chiral coupling is evident, though
somewhat diminished by the $\eta_{t, \ol t}$~cut. The forward--backward
asymmetry of~0.35 could be measured at the $5\sigma$ (statistical) level
with an integrated luminosity of $1\,\fb^{-1}$. For this luminosity, the
statistical errors on $d\sigma/d\cstar$ in six bins 0.30~units wide would
range from 20\% down to 10\%. This is one example of how useful it would be
to upgrade the Tevatron luminosity to~$\hl$.

The $\cstar$ distributions expected at the LHC for this $V_8$ are shown in
Fig.~14. In this example, the contribution of the $V_8$ is about 20\% of
the total and it is polluted by the $q \leftrightarrow \ol q$ ambiguity, so
that the rise in the cross section with $\cstar$ is invisible. The
asymmetry is only 2\%. This illustrates the dominance of $gg$ processes and
the uncertainty in determining the quark direction at small~$\tau$ in a
high--energy $pp$~collider. Essentially similar results were obtained for
the $\xi_t = 1/\xi_q$ case (see Table~3). We found that there is nothing to
be gained at the LHC by a looser $\eta_t$ cut, or by limiting the $\Mtt$
integration region to a narrow band about $\Mv$, or by selecting events
produced at large boost rapidity.

\newsec{Summary}

Top quarks, of all known elementary particles, are most intimately
connected to the physics of flavor and may provide keys to unlock its
mysteries. Thus, top--quark production at the Tevatron provides our most
incisive probe into flavor physics until the LHC turns on in the next
century. The invariant mass distributions that can be formed in top--quark
production appear to be the best means for distinguishing between standard
and nonstandard mechanisms.

The mean and RMS of the total invariant mass, $\Mtt$, provide an
independent measure of $m_t$ which should agree with the directly--measured
mass {\it if} production is governed by standard QCD. In QCD, the variance
$\Deltt$ is expected to be about 75~GeV. The total invariant mass can
reveal the presence of $\ttb$ resonances such as the top--color vectors
$V_8$~\hp,\topcolor\ and the technihadron $\eta_T$~\eekleta,\etatrefs. Such
resonances may easily double the $\ttb$ rate. It is worth noting that,
since the fraction of gluon--initiated processes rises fairly rapidly as
the machine energy, an upgrade of the Tevatron to $\sqrt{s} = 2\,\tev$ will
lead to quite different changes in the $V_8 \ra \ttb$ and $\eta_T \ra \ttb$
rates.\foot{I thank S.~Parke for emphasizing this point to me.} We find an
increase of about 50\% for $\qqb \ra V_8$, but almost 100\% for $gg \ra
\eta_T$. Subsystem invariant masses can be examined for alternative
explanations of the top--production data and for unconventional top decays.
In this regard, we emphasize that it may be dangerous to use the standard
QCD $\ttb$ production model to select top--candidate events. For example, a
resonance in $\ttb$ production may distort the summed scalar--$\et$ and
sphericity or aplanarity distributions of candidate events from their QCD
expectation.

The angular dependence of top--production may also provide valuable
information on the top--production mechanism. Although it is generally
expected that, for production near threshold, the angular distribution will
be isotropic, we have seen that chiral couplings can be detected if they
are present and comparable to the QCD amplitudes. The dominance of $\qqb$
annihilation in top--quark production processes at the Tevatron collider
gives it an advantage over the LHC for studying angular distributions. Two
distributions as different as those arising from the scalar--coupled
$\eta_T$ and the chiral--coupled $V_8$ may be distinguished with a data
sample of $1\,\fb^{-1}$. However, it is clear that the resolving power of
these distributions would benefit greatly from a significant upgrade of the
collider and its detectors so that samples of $\CO(10\,\fb^{-1})$ can be
collected.

In conclusion, we emphasize that the studies done here have all been at the
most naive parton level. We hope they will inspire the CDF and $\Dzero$
collaborations to undertake more realistic, detector--specific simulations
in the not--too--distant future.

\bigskip

I am indebted to Alessandra Caner, Sekhar Chivukula, Estia Eichten, John
Huth, Chris Quigg, Elizabeth Simmons, John Terning and Avi Yagil for
helpful conversations. This research was supported in part by the
Department of Energy under Grant~No.~DE--FG02--91ER40676.

\listrefs

\centerline{\vbox{\offinterlineskip
\hrule\hrule
\halign{&\vrule#&
  \strut\quad#\hfil\quad\cr
height4pt&\omit&&\omit&&\omit&&\omit&&\omit&&\omit&&\omit&\cr
&\hfill Run---Event \hfill&&\hfill$m_t$ \hfill&&\hfill
 $\Mtt$(before fit)\hfill&&\hfill
$\Mtt$(after fit) \hfill&&\hfill $\beta$(after fit)\hfill&&
\hfill $\cstar$\hfill &\cr
height4pt&\omit&&\omit&&\omit&&\omit&&\omit&&\omit&\cr
\noalign{\hrule}
height4pt&\omit&&\omit&&\omit&&\omit&&\omit&&\omit&&\omit&\cr
&40758--44414&&\hfill$172\pm 11$\hfill&&\hfill$523$\hfill
&&\hfill$526$\hfill&&
\hfill$0.757$\hfill&&\hfill $0.404$\hfill&\cr
height4pt&\omit&&\omit&&\omit&&\omit&&\omit&&\omit&&\omit&\cr
&43096--47223&&\hfill$166\pm 11$\hfill&&\hfill$533$\hfill
&&\hfill$511$\hfill&&
\hfill$0.760$\hfill&&\hfill $0.820$\hfill&\cr
height4pt&\omit&&\omit&&\omit&&\omit&&\omit&&\omit&&\omit&\cr
&43351--266423&&\hfill$158\pm 18$\hfill&&\hfill$440$\hfill
&&\hfill$460$\hfill&&
\hfill$0.727$\hfill&&\hfill $0.512$\hfill&\cr
height4pt&\omit&&\omit&&\omit&&\omit&&\omit&&\omit&&\omit&\cr
&45610--139604&&\hfill$180\pm 9$\hfill&&\hfill$338$\hfill
&&\hfill$366$\hfill&&
\hfill$0.180$\hfill&&\hfill $-0.0011$\hfill&\cr
height4pt&\omit&&\omit&&\omit&&\omit&&\omit&&\omit&&\omit&\cr
&45705--54765&&\hfill$188\pm 19$\hfill&&\hfill$440$\hfill
&&\hfill$431$\hfill&&
\hfill$0.489$\hfill&&\hfill $-0.348$\hfill&\cr
height4pt&\omit&&\omit&&\omit&&\omit&&\omit&&\omit&&\omit&\cr
&45879--123158&&\hfill$169\pm 10$\hfill&&\hfill$411$\hfill
&&\hfill$412$\hfill&&
\hfill$0.572$\hfill&&\hfill $-0.767$\hfill&\cr
height4pt&\omit&&\omit&&\omit&&\omit&&\omit&&\omit&&\omit&\cr
&45880--31838&&\hfill$132\pm 8$\hfill&&\hfill$384$\hfill
&&\hfill$365$\hfill&&
\hfill$0.691$\hfill&&\hfill $-0.682$\hfill&\cr
height4pt&\omit&&\omit&&\omit&&\omit&&\omit&&\omit&\cr}
\hrule\hrule}}
\medskip

\noindent Table 1. Best fit top--quark masses and kinematic characteristics
of the CDF experiment's $\ttb$ candidate events (from Ref.~\cdfpr). Masses
are in GeV. Transverse motion of the subprocess c.m. was neglected in
determining the top--quark velocity $\beta$ and scattering angle
$\theta^*$.

\bigskip\bigskip\bigskip\bigskip

\centerline{\vbox{\offinterlineskip
\hrule\hrule
\halign{&\vrule#&
  \strut\quad#\hfil\quad\cr\cr
height4pt&\omit&&\omit&&\omit&&\omit&&\omit&&\omit&&\omit&\cr\cr
&\hfill Model \hfill&&\hfill $\sigma(\ttb)$ \hfill&&\hfill
$\MMtt$  \hfill&&\hfill $m_t\ts(\MMtt)$\hfill&&\hfill $\RMStt$\hfill&&
\hfill $m_t\ts(\RMStt)$ \hfill&&\hfill $\Deltt$\hfill &\cr\cr
height4pt&\omit&&\omit&&\omit&&\omit&&\omit&&\omit&&\omit&\cr\cr
\noalign{\hrule\hrule}
height4pt&\omit&&\omit&&\omit&&\omit&&\omit&&\omit&&\omit&\cr\cr
&LO--QCD (EHLQ1)&&\hfill$5.13$\hfill&&\hfill$440$\hfill&&
\hfill$174$\hfill&&\hfill $447$\hfill&&\hfill $174$\hfill
&&\hfill $77$\hfill&\cr\cr
\noalign{\hrule}
height4pt&\omit&&\omit&&\omit&&\omit&&\omit&&\omit&&\omit&\cr\cr
&CDF data&&\hfill$13.9^{+6.1}_{-4.8}$\hfill&&\hfill$439$\hfill&&
\hfill$173$\hfill&&\hfill $443$\hfill&&\hfill $172$\hfill
&&\hfill $60$\hfill&\cr\cr
\noalign{\hrule}
height4pt&\omit&&\omit&&\omit&&\omit&&\omit&&\omit&&\omit&\cr\cr
&$M_{V_8^-} = 450$&&\hfill$13.3$\hfill&&\hfill$431$\hfill&&
\hfill$170$\hfill&&\hfill $433$\hfill&&\hfill $168$\hfill
&&\hfill $46$\hfill&\cr\cr
height4pt&\omit&&\omit&&\omit&&\omit&&\omit&&\omit&&\omit&\cr\cr
&$M_{V_8^+} = 450$&&\hfill$11.0$\hfill&&\hfill$465$\hfill&&
\hfill$185$\hfill&&\hfill $469$\hfill&&\hfill $184$\hfill
&&\hfill $58$\hfill&\cr\cr
\noalign{\hrule}
height4pt&\omit&&\omit&&\omit&&\omit&&\omit&&\omit&&\omit&\cr\cr
&$M_{V_8^-} = 475$&&\hfill$14.9$\hfill&&\hfill$440$\hfill&&
\hfill$174$\hfill&&\hfill $444$\hfill&&\hfill $173$\hfill
&&\hfill $53$\hfill&\cr\cr
height4pt&\omit&&\omit&&\omit&&\omit&&\omit&&\omit&&\omit&\cr\cr
&$M_{V_8^+} = 475$&&\hfill$10.8$\hfill&&\hfill$482$\hfill&&
\hfill$193$\hfill&&\hfill $487$\hfill&&\hfill $192$\hfill
&&\hfill $67$\hfill&\cr\cr
\noalign{\hrule}
height4pt&\omit&&\omit&&\omit&&\omit&&\omit&&\omit&&\omit&\cr\cr
&$\Mh = 450$&&\hfill$13.5$\hfill&&\hfill$432$\hfill&&
\hfill$171$\hfill&&\hfill $435$\hfill&&\hfill $169$\hfill
&&\hfill $52$\hfill&\cr\cr
height4pt&\omit&&\omit&&\omit&&\omit&&\omit&&\omit&&\omit&\cr\cr
&$\Mh = 475$&&\hfill$11.4$\hfill&&\hfill$442$\hfill&&
\hfill$175$\hfill&&\hfill $446$\hfill&&\hfill $174$\hfill
&&\hfill $55$\hfill&\cr\cr
\noalign{\hrule}
height4pt&\omit&&\omit&&\omit&&\omit&&\omit&&\omit&&\omit&\cr\cr
&$t_s(160)$ $t(175)$&&\hfill$13.2$\hfill&&\hfill$421$\hfill&&
\hfill$166$\hfill&&\hfill $428$\hfill&&\hfill $166$\hfill
&&\hfill $77$\hfill&\cr\cr
height4pt&\omit&&\omit&&\omit&&\omit&&\omit&&\omit&&\omit&\cr\cr
&$t_s(165)$ $t(190)$&&\hfill$10.0$\hfill&&\hfill$437$\hfill&&
\hfill$173$\hfill&&\hfill $444$\hfill&&\hfill $173$\hfill
&&\hfill $77$\hfill&\cr\cr
height4pt&\omit&&\omit&&\omit&&\omit&&\omit&&\omit&&\omit&\cr\cr}
\hrule\hrule}}
\medskip

\medskip

\noindent Table 2. $\ppb \ra \ttb$ cross sections (in pb) at $\sqrt{s} =
1800\,\gev$ and their kinematic characteristics for lowest--order QCD, CDF
data~\cdfpr, and the three nonstandard production models with parameters
described in the text. Cross sections have been multiplied by 1.62.

\vfil\eject

\centerline{\vbox{\offinterlineskip
\hrule\hrule
\halign{&\vrule#&
  \strut\quad#\hfil\quad\cr
height4pt&\omit&&\omit&&\omit&&\omit&&\omit&&\omit&&\omit&\cr
&\hfill Model \hfill&&\hfill $\Mtt$ range \hfill&&\hfill
 Collider\hfill&&\hfill
$\stt$ \hfill&&\hfill $\sigma_F$ \hfill&&
\hfill $\sigma_B$ \hfill&& \hfill $A_{FB}$ \hfill &\cr
height4pt&\omit&&\omit&&\omit&&\omit&&\omit&&\omit&&\omit&\cr
\noalign{\hrule}
height4pt&\omit&&\omit&&\omit&&\omit&&\omit&&\omit&&\omit&\cr
&\hfill $\eta_T$ \hfill&&\hfill $430-470$ \hfill&&\hfill TEV\hfill
&&\hfill$4.82$\hfill&&
\hfill$2.41$\hfill&&\hfill $2.41$\hfill&&\hfill 0$$ \hfill &\cr
height4pt&\omit&&\omit&&\omit&&\omit&&\omit&&\omit&&\omit&\cr
&\hfill $\eta_T$ \hfill&&\hfill $430-470$ \hfill&&\hfill LHC\hfill
&&\hfill$4360$\hfill&&
\hfill$2180$\hfill&&\hfill $2180$\hfill&&\hfill $0$ \hfill &\cr
height4pt&\omit&&\omit&&\omit&&\omit&&\omit&&\omit&&\omit&\cr
&$V_{8^-}$ &&\hfill $400-500$ \hfill&&\hfill TEV \hfill
&&\hfill$8.14$\hfill&&
\hfill$5.48$\hfill&&\hfill $2.66$\hfill&&\hfill $0.35$ \hfill &\cr
height4pt&\omit&&\omit&&\omit&&\omit&&\omit&&\omit&&\omit&\cr
&$V_{8^-}$ &&\hfill $400-500$ \hfill&&\hfill LHC\hfill
&&\hfill$285$\hfill&&
\hfill$145$\hfill&&\hfill $140$\hfill&&\hfill $0.017$ \hfill &\cr
height4pt&\omit&&\omit&&\omit&&\omit&&\omit&&\omit&&\omit&\cr
&$V_{8^+}$ &&\hfill $425-525$ \hfill&&\hfill TEV \hfill
&&\hfill$5.93$\hfill&&
\hfill$4.21$\hfill&&\hfill $1.72$\hfill&&\hfill $0.42$ \hfill &\cr
height4pt&\omit&&\omit&&\omit&&\omit&&\omit&&\omit&&\omit&\cr
&$V_{8^+}$ &&\hfill $425-525$ \hfill&&\hfill LHC \hfill
&&\hfill$255$\hfill&&
\hfill$130$\hfill&&\hfill $125$\hfill&&\hfill $0.021$ \hfill &\cr
height4pt&\omit&&\omit&&\omit&&\omit&&\omit&&\omit&&\omit&\cr}
\hrule\hrule}}
\medskip

\noindent Table 3. Angular dependences of $\ttb$ production in the $\eta_T$
and $V_8$ resonance models with parameters described in the text. Top
quarks are produced with pseudorapidity $|\eta| < 2.0$ and cross sections
(in pb) have been multiplied by 1.62.

\vfil\eject

\centerline{\bf Figure Captions}
\bigskip

\item{[1]} The $\ttb$ invariant mass distributions, in $\ppb$ collisions at
$\ecm = 1800\,\gev$, for $m_t = 100-220\,\gev$ in $20\,\gev$ increments.
EHLQ1 distribution functions were used and the cross sections were
multiplied by~1.62 as explained in the text. No rapidity cut is applied.

\medskip

\item{[2]} The mean (solid) and root--mean--square (dashed) $\ttb$
invariant mass, as a function of $m_t$, for $\ppb \ra \ttb$ at $\sqrt{s} =
1800\,\gev$. Lowest--order QCD cross sections (Fig.~1) were used.

\medskip

\item{[3]} The $\ttb$ invariant mass distribution in the presence of a
$V_8$, in $\ppb$ collisions at $\ecm = 1800\,\gev$, for $m_t = 175\,\gev$
and $\Mv = 450\,\gev$, $\xi_t = \xi_b = -1/\xi_q = \sqrt{40/3}$. The QCD
(dotted curve) and the total (solid) rates have been multiplied by 1.62 as
explained in the text. No rapidity cut is applied to the top quarks.

\medskip

\item{[4]} The $\ttb$ invariant mass distribution in the presence of a
$V_8$, in $\ppb$ collisions at $\ecm = 1800\,\gev$. The parameters and
curves are as in Fig.~3 except that $\xi_t = \xi_b = 1/\xi_q = \sqrt{40/3}$.

\medskip

\item{[5]} The $\ttb$ invariant mass distribution in the presence of a
$V_8$, in $\ppb$ collisions at $\ecm = 1800\,\gev$, for $m_t = 175\,\gev$
and $\Mv = 475\,\gev$, $\xi_t = \xi_b = -1/\xi_q = \sqrt{40/3}$. The curves
are labeled as in Fig.~3.
\medskip

\item{[6]} The $\ttb$ invariant mass distribution in the presence of a
$V_8$, in $\ppb$ collisions at $\ecm = 1800\,\gev$. The parameters and
curves are as in Fig.~5 except that $\xi_t = \xi_b = 1/\xi_q = \sqrt{40/3}$.

\medskip

\item{[7]} The $\ttb$ invariant mass distribution in the presence of an
$\eta_T$, in $\ppb$ collisions at $\ecm = 1800\,\gev$, for $m_t =
175\,\gev$ and $\Mh = 450\,\gev$, $F_Q = 30\,\gev$ and $C_t = -1/3$. The
QCD (dotted curve), $\eta_T \ra \ttb$ and its interference with the QCD
amplitude (dashed), and total (solid) rates have been multiplied by 1.62
as explained in the text. No rapidity cut is applied to the top quarks.

\medskip

\item{[8]} The $\ttb$ invariant mass distribution in the presence of an
$\eta_T$, in $\ppb$ collisions at $\ecm = 1800\,\gev$. The parameters and
curves are as in Fiig.~7 except that $\Mh = 475\,\gev$.

\medskip

\item{[9]} The effective $\ttb$ mass distribution for $\ppb \ra \ttb$
(dotted) and $t_s \ol t_s$ (dashed) at $\sqrt{s} = 1800\,\gev$; $m_{t_s} =
160\,\gev$ and $m_t = 175\,\gev$. The solid corve is the sum of the two
mass distributions.

\medskip

\item{[10]} The effective $\ttb$ mass distribution for $\ppb \ra \ttb$ and
$t_s \ol t_s$ at $\sqrt{s} = 1800\,\gev$; $m_{t_s} = 165\,\gev$ and $m_t =
190\,\gev$. Curves are labeled as in Fig.~9.

\medskip

\item{[11]} The $\cstar$ distribution for $\ppb \ra \ttb$ at $\ecm =
1800\,\gev$ in the presence of a 450~GeV $\eta_T$ with parameters as in
Fig.~7; $430 < \Mtt < 470\,\gev$. The components are standard QCD $gg \ra
\ttb$ (dot-dash), $\qqb \ra \ttb$ (long dashes), total QCD (dots), $gg \ra
\eta_T \ra \ttb$ and interference with QCD (short dashes), and the total
$d\sigma/\cstar$ (solid). EHLQ1 distribution functions were used and
all cross sections were multiplied by 1.62. The top quarks are required to
have pseudorapidity $|\eta| < 2.0$.

\medskip

\item{[12]} The $\cstar$ distribution for $pp \ra \ttb$ at $\ecm =
15\,\tev$ in the presence of a 450~GeV $\eta_T$ with parameters as in
Fig.~11; $430 < \Mtt < 470\,\gev$. The curves are labeled as in Fig.~11.

\medskip

\item{[13]} The $\cstar$ distribution for $\ppb \ra \ttb$ at $\ecm =
1800\,\gev$ in the presence of a 475~GeV $V_8$ with parameters as in
Fig.~5; $400 < \Mtt < 500\,\gev$. The components are standard QCD $gg \ra
\ttb$ (dot-dash), $\qqb \ra \ttb$ (long dashes), total QCD (dots), $\qqb
\ra V_8 \ra \ttb$ and interference with QCD (short dashes), and the total
$d\sigma/\cstar$ (solid). EHLQ1 distribution functions were used and
all cross sections were multiplied by 1.62. The top quarks are required to
have pseudorapidity $|\eta| < 2.0$.

\medskip

\item{[14]} The $\cstar$ distribution for $pp \ra \ttb$ at $\ecm = 15\,\tev$
in the presence of a 475~GeV $V_8$ with parameters as in Fig.~13; $400 <
\Mtt < 500\,\gev$. The curves labeled as in Fig.~13.

\vfil\eject

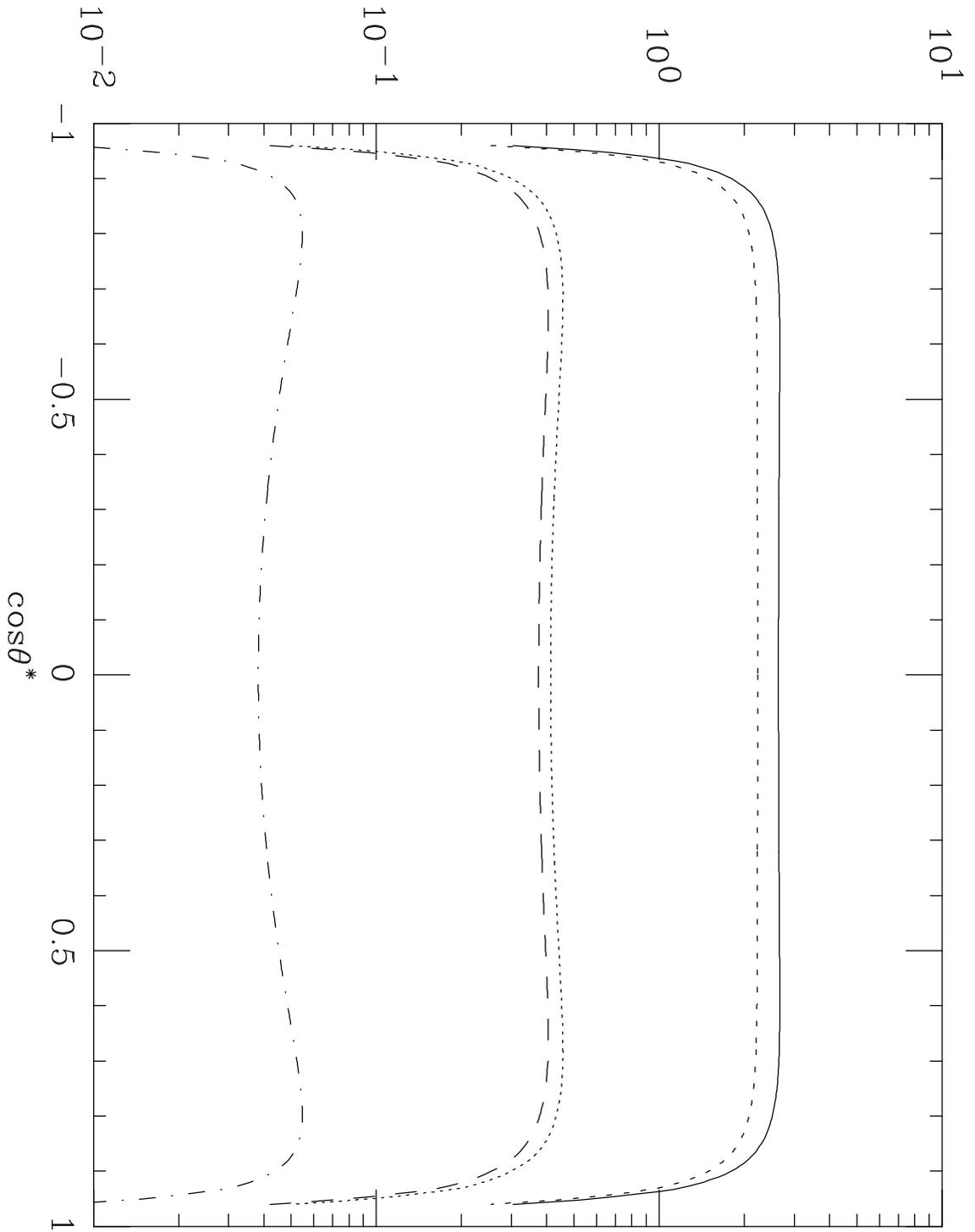
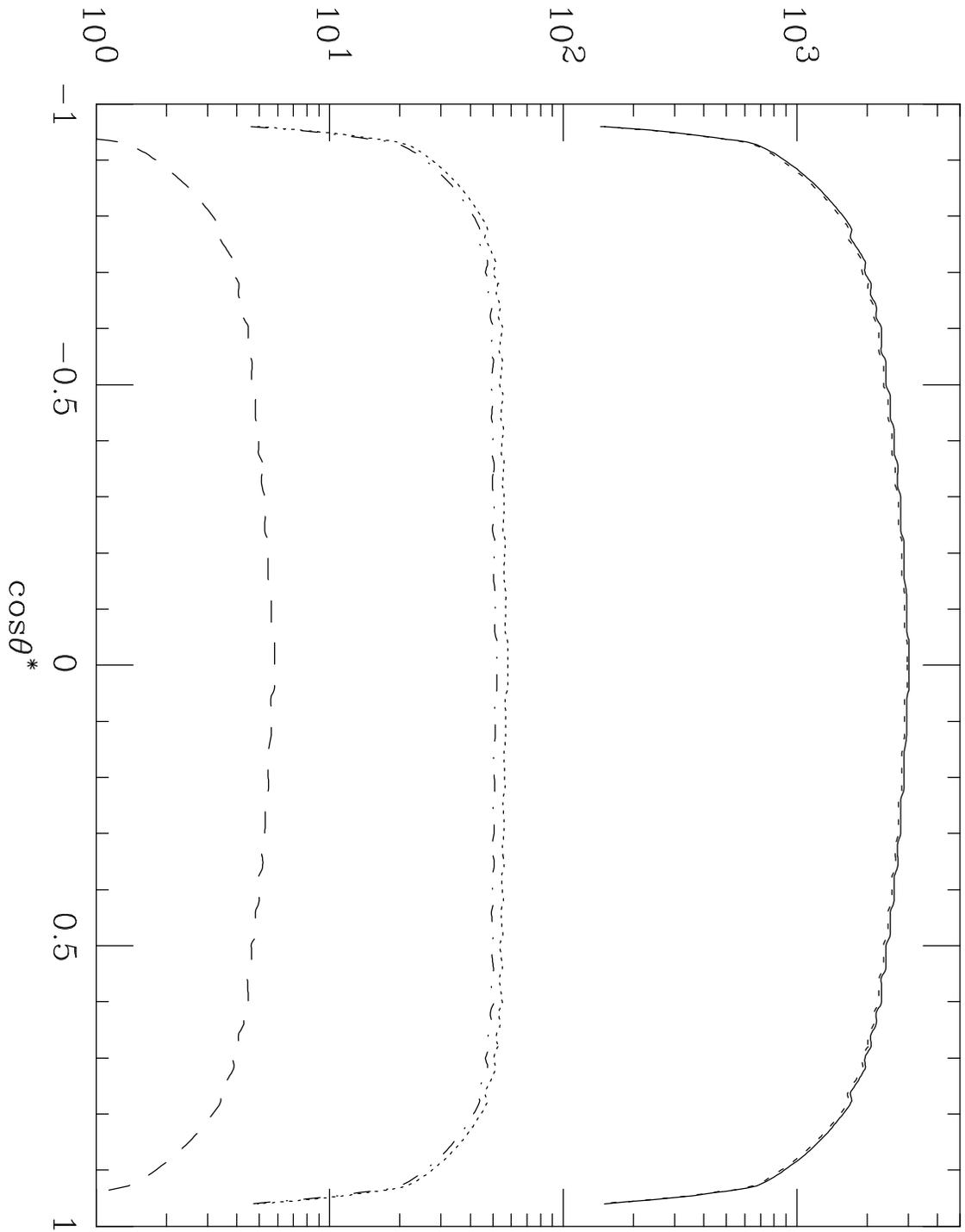
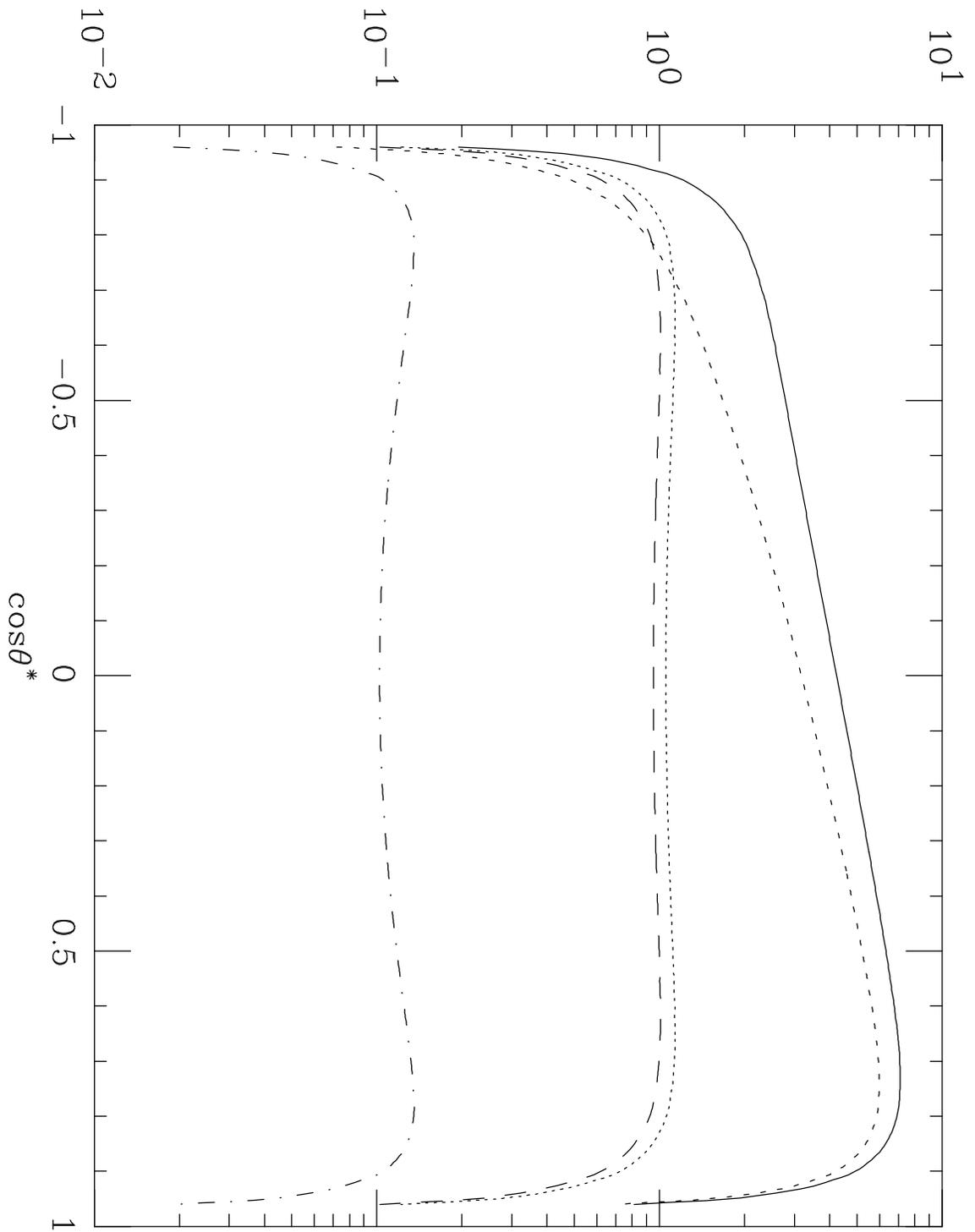
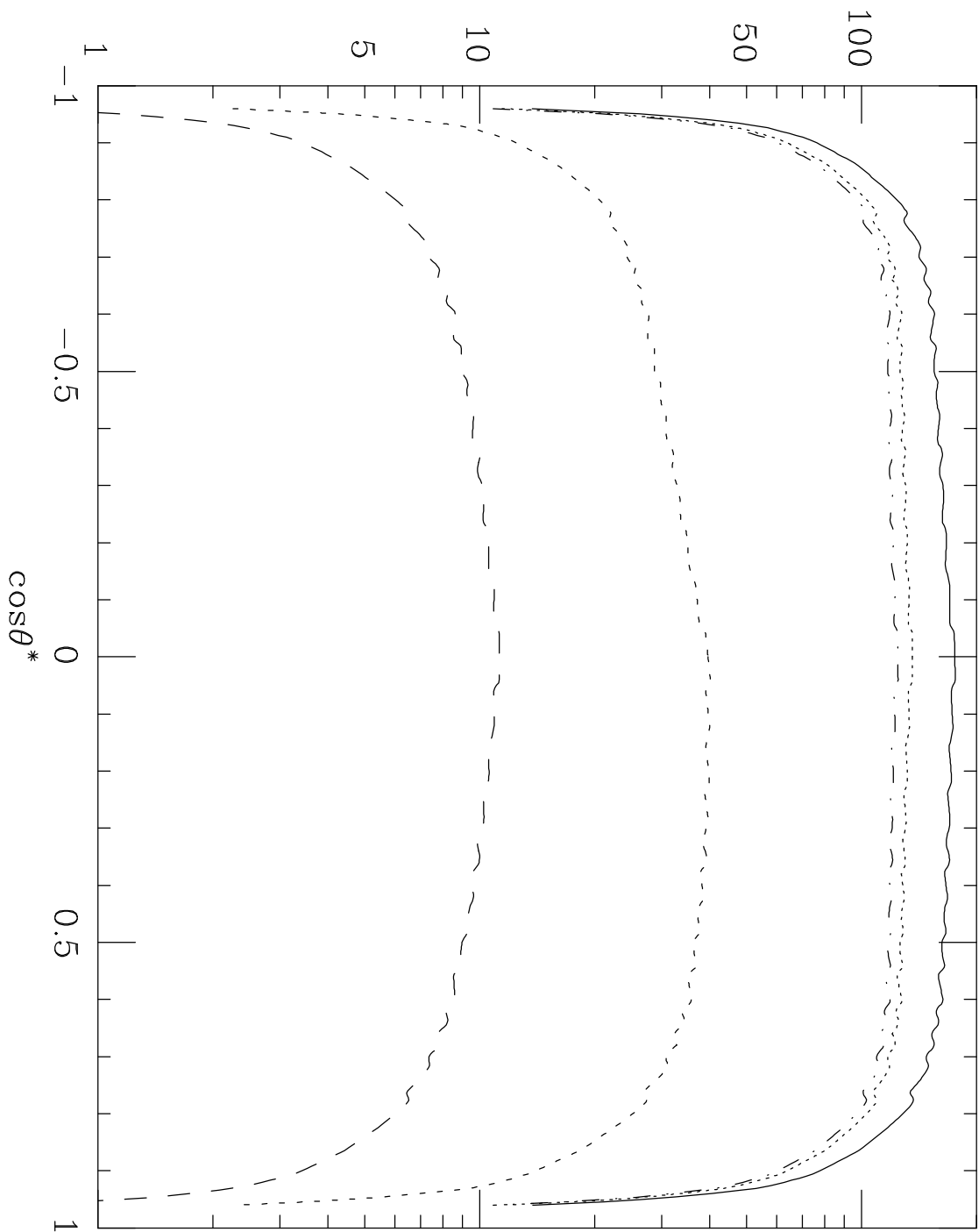
\bye